\documentclass[journal,twoside]{IEEEtran}

\usepackage{defpaper}
\usepackage{epsfig}
\usepackage{latexsym}

\newcommand{\filefig}[4]{
  \begin{figure}[htb]
  \vskip 1pt
    \begin{center}
      \setlength{\epsfxsize}{#4}
      \leavevmode
      \epsfbox {#2}
      \caption{\protect {#1}}
      \label {#3}
   \end{center}
\end{figure}}

      {
      \begin{minipage}
      {0.462\textwidth}\medskip\noindent
       \scriptsize\raggedright\sf\noindent
       {\bf Input:} #2\\
       {\bf Output:} #3\\
       \rule[1pt]{1.0\textwidth}{0.3pt}%
       \setlength{\leftmargini}{0em}%
       \setlength{\leftmarginii}{1.5em}%
       \setlength{\leftmarginiii}{1.5em}%
       \setlength{\leftmarginiv}{1.5em}%
       \setlength{\leftmarginv}{1.5em}%
       \begin{description}}
      {\end{description}\medskip
       \end{minipage}
       }

\long\def\comment#1{}

\long\def\symbolfootnote[#1]#2{\begingroup%
\def\thefootnote{\fnsymbol{footnote}}\footnote[#1]{#2}\endgroup}

\ifCLASSINFOpdf
\else
\fi

\begin{document}
%
\title{\emph{vFlow}: A GUI-Based Tool for Building Batch Applications for Cloud Computing}

\author{\begin{tabular}[t]{c@{\extracolsep{1em}}c@{\extracolsep{1em}}c}
           Kamal A. Ahmat & Hassan Gobjuka\\
\it        Department of Information Technology & \it Verizon \\
\it        City University of New York & \it 919 Hidden Ridge \\
\it        New York, NY 11101 & \it Irving, TX 75038 \\
\it        {\small\tt kamal.ahmat@live.lagcc.cuny.edu} & \it {\small\tt hasan.gobjuka@verizon.com}
\end{tabular}}

\markboth{IEEE INFOCOM Demo Session 2011}%
{Ahmat  \MakeLowercase{and} Gobjuka: A GUI-Based Tool for Building Batch Applications for Cloud Computing}

\maketitle

\begin{abstract}
In this paper we introduce \emph{vFlow} -– A framework for rapid designing of batch processing applications for Cloud Computing environment. \emph{vFlow} batch processing system extracts tasks from the \emph{vPlans} diagrams, systematically captures the dynamics in batch application management tasks, and translates them to Cloud environment API, named \emph{vDocuments}, that can be used to execute batch processing applications. \emph{vDocuments} do not only enable the complete execution of low-level configuration management tasks, but also allow the construction of more sophisticated tasks, while imposing additional reasoning logic to realize batch application management objectives in Cloud environments. We present the design of the \emph{vFlow} framework and illustrate its utility by presenting the implementation of several sophisticated operational tasks.

\end{abstract}

\IEEEpeerreviewmaketitle

\section{Introduction}
The rapid growth of demand on modern resource-intensive distributed enterprise applications in virtualized environments has led to a significant need to advanced tools that are capable of automating the management of such applications. Even though the research community (e.g. \cite{warn}) and industry (e.g. \cite{pathak}) have been very actively exploring methods for automating resource management in grid and batch processing environments, these works mainly focus on the backend part, and consequently, there is a little work done to allow the automatic creation of batch processing applications in Cloud environments, which can substantially reduce the human factor errors and drastically improve the dynamically automated resource management process.

In this paper we present our novel state-of-the-art framework, \emph{vFlow}, for rapidly building automated batch applications in virtualized environments based on Petri Net model \cite{PetriNets}. \emph{vFlow} allows Cloud users to automate application execution and management process by building distributed batch processing application flow plans known as \emph{vPlans}. \emph{vPlans} can be rapidly created through \emph{vFlow}'s GUI-based editor, and also can be modified through the textual editor. The editor systematically captures the dynamics in distributed data-intensive complex batch processing application management tasks, validates them and translates them to formal batch application execution plan that can be executed in Cloud environments such as Amazon EC2\footnote{http://aws.amazon.com/ec2/} and GoGrid\footnote{http://www.gogrid.com}. Our tool supports user interactions with large-scale Cloud environments through its interfaces without the need for plug-ins or special-purpose runtime systems and generates validated ready-to-run batch processing application execution plans.

\section{Presentation}
\label{model}

\subsection{Background}

Petri Nets \cite{PetriNets} are flow-chart-like diagrams that were first introduced by Carl Adam Petri in 1962 as a formal method for describing the concurrency and synchronization present in distributed systems and to determine the correctness of these systems. The main elements of Petri Nets are as follows:

\begin{itemize}
  \item \textbf{Places}: These circular nodes represent conditions or objects such as program variables. A place may contain none, one, or multiple \emph{Tokens};
  \item \textbf{Tokens}: These Black dots represent the specific state or the condition or object. The places that have tokens are represent the current state of the described system;
  \item \textbf{Transitions}: Transitions, which are represented as rectangles, model system activities which result in a change in the current system status. Transitions can be method calls, for instance. A transition executes if all Places that have input \emph{Arcs} to the transition have tokens;
  \item \textbf{Arcs}: These arrows connect between pairs of places and transitions. Arcs show the relative order and dependency of between various places and transitions and thus indicate which objects and conditions are changed by which activity.
\end{itemize}

For more information about Petri Nets and its usages in representing application flow, we refer the reader to \cite{PetriNets}.

\subsection{Framework}
\label{framework}

The framework consists of the main five loosely coupled components: the \emph{vPlan} GUI Editor; the \emph{vPlan} Textual Editor;  the \emph{vPlan} Validator; the Repository (Execution \emph{vPlan} Library); and the API Generator.

\filefig{The architecture used in \emph{vFlow}.}
{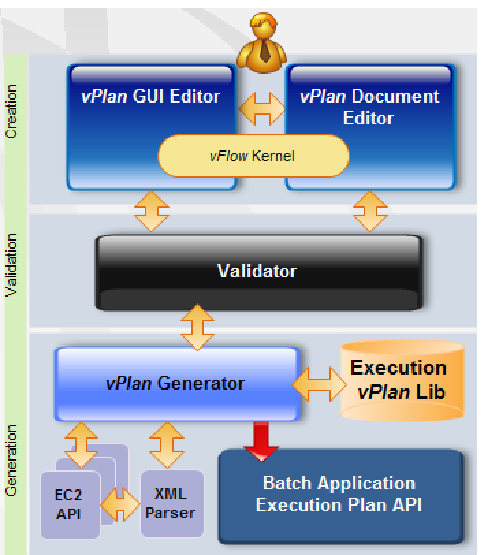}    
{arch}      
{2.3in}        

\begin{itemize}
  \item \textbf{\emph{vPlan} GUI Editor:} The editor is a central component in \emph{vFlow} framework. The editor is the GUI-based user interface component that allows the user to create a distributed batch application flow model process based on Petri Net model techniques. The user interface is similar to a flow chart where the user could put circles, squares and triangles, then connect them together. Although this seems simple, the diagrams can become quite complex based on the Petri Net model controls of these symbols. The editor also allows building complex hierarchical application execution plans. Within each circle in the \emph{vPlan}, code can be running from Java or Ruby sources, or the user can choose to generate application execution plan document (Web Service API) by integrating the execution plan with a specific Cloud vendor's API. The editor captures advanced features such as annotations and interrupts when an event occurs in the application running in Cloud.
  \item \textbf{\emph{vPlan} Textual Editor:} The textual editor allows the user to modify the initial application execution plan generated by GUI Editor. This includes for instance adding comments to the plan document. The modifications made to the plan document through the Textual Editor are reflected in real-time to the GUI editor, and vice versa.
  \item \textbf{\emph{vPlan} Validator:} This component is responsible of the verification and validation of the \emph{vPlan} structure created by the user. It mainly verifies the following features: \textbf{(1)\emph{Well-Formedness}:} There is exactly one main \emph{vPlan} per application, which is connected, and has one start and one end places; \textbf{(2)\emph{Reachability}:} There must be at least one path from the start place to the end place passing through every other place and transition in each \emph{vPlan}; and \textbf{(3)\emph{Well-Structureness}:} If the \emph{vPlan} is hierarchical, then each lower level \emph{vPlan} module referred by meta-place in the higher level \emph{vPlan} is designed and exists. Furthermore, it assures that annotations represent valid arguments with respect to the corresponding transitions.
  \item \textbf{Repository:} This is the code repository to store all modules and related scripts. Each transition represents a call to a module or script from the repository and repository content can be reused.
  \item \textbf{API Generator:} Generates the final batch application execution plan in form of Web Service API. Compared to traditional scripts that at best automate the generation and modification of configuration on target application or device, a \emph{vPlan} API additionally encapsulates the logical reasoning that guides the workflow of the application management task. This capability enables full automation of batch processing application tasks, minimizing human involvement.
\end{itemize}

\filefig{The \emph{vPlan} default editor interface.}
{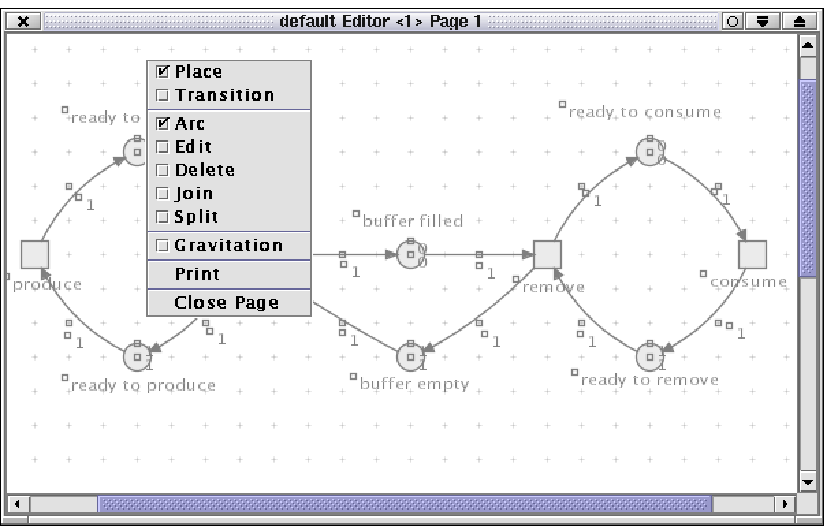}    
{arch}      
{3.4in}        

\section{\emph{vFlow} Demonstration}

The demo software is a complete implementation of the \emph{vFlow} framework, featuring the components described in Section \ref{framework}. In our demo we plan on showing various abilities of \emph{vFlow}. The user first creates a batch application execution plan using the GUI editor and assigns each transition with one of the functions (i.e. tasks) written in advance and stored in the repository. The users alternatively can develop their own functions using Java or Ruby. Once the \emph{vPlan} is validated, the API Generator generates ready-to-run \emph{vDocument} (i.e. Amazon EC2 Web Service API), and the editor allows the user to go through, step-by-step, to test/debug the application or just a specific module. The user can also edit the plan via the \emph{vPlan} Textual Editor. The developed \emph{vPlan} can be run repeatedly based on specific configurations (e.g. once every 24 hours at 12:00 AM) and it runs in the background, so that it doesn't affect the performance of other critical applications.

\end{document}